\documentclass[epsfig,12pt]{article}
\usepackage{epsfig}
\usepackage{graphicx}
\usepackage{array}

\newcommand{\beq}{\begin{equation}}   
\newcommand{\eeq}{\end{equation}}
\newcommand{\beqn}{\begin{eqnarray}}   
\newcommand{\eeqn}{\end{eqnarray}}

\newcommand{\gsim}{\lower.7ex\hbox{$
\;\stackrel{\textstyle>}{\sim}\;$}}
\newcommand{\lsim}{\lower.7ex\hbox{$
\;\stackrel{\textstyle<}{\sim}\;$}}
\begin{document}

\begin{titlepage}

\begin{flushright}
FTPI-MINN-13/34, UMN-TH-3305/13\\
September 17/2013
\end{flushright}

\vspace{0.7cm}

\begin{center}
{  \large \bf  Degeneracy between Abelian and Non-Abelian Strings}
\end{center}
\vspace{0.6cm}

\begin{center}
 {\large 
  Sergei Monin,$^{a}$   M. Shifman,$^b$}
\end {center}

\vspace{3mm}
 
\begin{center}
$^a${\em Department of Physics, University of Minnesota,
Minneapolis, MN 55455, USA} \\[1mm]
$^b${\em William I. Fine Theoretical Physics Institute, University of Minnesota,
Minneapolis, MN 55455, USA}
\end {center}

\vspace{2cm}

\begin{center}
{\large\bf Abstract}
\end{center}

In a model that supports both Abelian (Abrikosov-Nielsen-Olesen) and non-Abelian strings
we analyze the parameter space to find examples in which these strings not only coexist but are degenerate in tension.
We prove that both  solutions are locally stable, i.e there are no negative modes in the 
string background. The tension degeneracy is achieved at the classical level
and is expected to be lifted by quantum corrections. The set up of the model, analogous to that of Witten's superconducting cosmic strings, had  been extended to include non-Abelian strings previously \cite {S}.

\hspace{0.3cm}

\end{titlepage}

\newpage

\section{Introduction}

This paper can be viewed as a logical continuation of
\cite{S,Sh,MSY}. Topologically stable non-Abelian strings\footnote{I.e. those strings that have non-Abelian moduli fields on the string world sheet.} which were discovered \cite{AHDT} 
in 2003 significantly extended the class of the (local) stringy solitons which essentially had been limited previously
to the Abrikosov-Nielsen-Olesen (ANO) strings \cite{ANO}. A natural question arises as to whether the 
ANO (i.e. Abelian) and 
non-Abelian strings can coexist in one and the same model, both being locally stable, and if yes, whether their tensions can be degenerate. The exact answer to the second question can be given only in supersymmetric models provided that both strings are BPS-saturated \cite{BPS}, with one and the same central charge.

Deferring this task for the future here we will explore a simple non-supersymmetric model \cite{S} which extends Witten's superconducting cosmic strings \cite{W} to find whether or not (classically) degenerate Abelian and non-Abelian strings are simultaneously supported in this model for at least some values of parameters.  
We will analyze the parameter space to find examples of degenerate strings which are
locally stable, i.e there are no negative modes in the 
string background.

We mainly follow the recent publication \cite{MSY} to (numerically) construct  profile functions with
zero and non-zero values of the triplet field $\chi$, i.e.  Abelian vs. non-Abelian.  To justify the quasiclassical approximation we assume weak coupling in the bulk. First, to normalize our calculation,  we determine the profile functions corresponding to the Abrikosov-Nielsen-Olesen string and find its tension. Next, we find the string solution with non-zero $\chi$. We show that with the appropriate choice of the parameters the two strings are degenerate in tension at the classical level
(within the accuracy of our numerical calculations). We also investigate stability of the strings.

\section{Formulation of the problem}
\vspace{2mm}

The model in which we will analyze string-like solitons is described by the Lagrangian
\beq
{\cal L} = {\cal L}_{0} +{\cal L}_\chi
\label{odin}
\eeq
where
 \beqn
{\cal L}_{0} &=& -\frac{1}{4e^2}F_{\mu\nu}^2 + \left| {\mathcal D}^\mu\phi\right|^2
  -V(\phi )\, ,
  \nonumber\\[2mm]
  {\mathcal D}_\mu\phi &=& (\partial_\mu -iA_\mu )\phi\,,
\nonumber\\[2mm]
  V &=& \lambda \left(|\phi |^2 -v^2
\right)^2\,,
  \label{tpi16}
\eeqn
and
\beqn
{\cal L}_\chi &=& \partial_\mu \chi^i \, \partial^\mu \chi^i  - U(\chi, \phi)\,,
\label{14}\\[2mm]
U &=&  \gamma\left[\left(-\mu^2 +|\phi |^2
\right)\chi^i \chi^i + \beta \left( \chi^i \chi^i\right)^2\right],
\label{15p}
\eeqn
with $i=1,2,3$. We will assume that $\lambda, \beta ,\gamma  > 0$ and $v^2 > \mu^2$. This model has the U(1) gauge symmetry, while the $\chi$ sector is O(3) symmetric. The $\chi$ fields are real.

The potential $V(\phi)$ ensures the Higgsing of the $U(1)$ photon. The complex scalar field $\phi$ develops a 
nonvanishing vacuum expectation value (VEV), 
$$|\phi|=v\,.$$
 As a result of the Higgs mechanism the phase of the comlex field is eaten up and becomes photon's longitudinal component. The photon   mass is
\beq
m^2_A=2e^2v^2\,.
\label{h6} 
\eeq
The physical Higgs excitation obviously has the  mass
\beq
m^2_{\phi}=4\lambda v^2\, .
\label{h7} 
\eeq
As can be seen from (\ref{15p}), the triplet field $\chi^i$ does {\em not} condense in the vacuum. Its mass is
\beq
m^2_{\chi}=\gamma\left(v^2-\mu^2\right).
\label{h8}
\eeq

For what follows it is convenient to introduce three auxiliary dimensionless parameters:
\beq
a=\frac{m^2_A}{m^2_{\phi}}\equiv\frac{e^2}{2\lambda}\,,\quad b=\frac{m^2_{\chi}}{m^2_\phi} \equiv \frac{\gamma}{4\lambda}\,\frac{c-1}{c}\,,\quad c=\frac{v^2}{\mu^2}\,.
\eeq

As was discussed in \cite{MSY},  a constraint  on the parameters of the Lagrangian exists from the requirement of vacuum stability, namely,
\beq
\beta \geq \beta_*\equiv\frac{b}{c(c-1)}\,.
\eeq

Other than that the parameters $e$, $\lambda$, $\beta$, $\mu$, and $v$ can be chosen at will. We will assume them to be chosen in such a way that the bulk model is weakly coupled and, hence, the quasiclassical approximation is applicable.

\vspace{0.5cm}
\newcolumntype{C}{ >{\centering\arraybackslash} p{3cm} }
\newcolumntype{D}{ >{\centering\arraybackslash} p{3cm} }
\begin{table}[h!]
\centering
\renewcommand\arraystretch{3.5}
\begin{tabular}{|C|D|}
\hline
$\beta$ & $\displaystyle\frac{\tilde\lambda}{\gamma}$ \\\hline
$a$ & $\displaystyle\frac{m_A^2}{m_{\phi}^2}$ \\\hline
$b$ & $\displaystyle\frac{m_{\chi}^2}{m_{\phi}^2}$ \\\hline
$\displaystyle\frac{v^2}{\mu^2}\equiv c$ & $\displaystyle \left( 1-\frac{4\lambda}{\gamma}\frac{m_{\chi}^2}{m_{\phi}^2}\right)^{-1}$ \\\hline
\end{tabular}
\caption{\small Parameters in (\ref{tpi16}), (\ref{15p}) in terms of the particle masses and the coefficients in front of 
the quartic terms $\phi^4$, $\chi^4$, and $\phi^2\chi^2$  ($\lambda$, $\tilde\lambda$, and $\gamma$, respectively).}
\label{t1}
\end{table}

Table \ref{t1} shows how the parameters in (\ref{tpi16}), (\ref{15p}) and $a,b,c$ are related  to the masses  
of the particles involved and the 
 coefficients in front of the quartic terms $\phi^4$, $\chi^4$ and $\phi^2\chi^2$. One can view the latter as ``physical" parameters of the problem at hand. 
 
 \section{Key observation}
 
 Since the Lagrangian (\ref{odin}) -- (\ref{15p}) does not contain terms linear in $\chi$,
 classical equations of motion always have a solution with $\chi \equiv 0$. If we set $\chi =0$ from the beginning,
our model reduces to the standard scalar QED, which supports the standard ANO string. Needless to say that
the latter is topologically stable.

Thus, our task  is, in addition to the ANO string, to find a solution with $\chi \neq 0$ in the string core.
Then we must check that this solution is (classically) stable under small perturbations with respect to
desintegration into the ANO string plus $\chi$ quanta. 
 
 \section{Anz\"atze and classical equations}

The basic steps of the appropriate construction were discussed in \cite{MSY,S,SA}. Here we will only outline its main points. Let us assume that the string lies along the $z$ axis, and  introduce a dimensionless radius in the perpendicular $\{x,y\}$ plane,
\beq
\rho = m_\phi \, \sqrt {x^2 + y^2 }\,.
\eeq

If we look for the  topologically stable strings, we must make the $\phi$ field  wind around the
string axis. For simplicity we will assume the minimal (unit) winding. The appropriate ans\"{a}tze are
\beq
 A_0=0\,,\quad A_i=-\varepsilon_{ij}\frac{x_j}{r^2}\left(\rule{0mm}{4mm} 1-f(r)\right)\,,\quad
\phi=v\varphi(\rho )e^{i\alpha}\,,
\label{aaa}
\eeq
where $\alpha$ is the polar angle in the perpendicular plane. The boundary conditions are obvious, 
\beq
f(\infty)=0\,,\qquad f(0)=1\,;\qquad\,
 \varphi(\infty)=1 \,,\quad \varphi(0)=0\,.
 \label{bc1}
\eeq

From Eq.~(\ref{15p}) it is clear that the vanishing value of the $\phi$ field in the core of the string may or may not generate a nonvanishing  value of the $\chi^i$ field. This is a dynamical issue. The $\chi = 0$ solution always exists and is stable, while the $\chi \neq 0$ solution may not exist or, if exist, may be classically unstable (i.e. the maximum of theenergy functional rather than its minimum). If the $\chi \neq 0$ solution exists its natural normalization is
\beq
(\chi^i\chi^i)_{\rm core} \approx \frac{\mu^2}{2\beta}\,.
\label{core}
\eeq
The existence of a stable solution with $\chi \neq 0$ in the  core 
implies that the O(3) symmetry is broken on the string. The appropriate ansatz for $\chi$ is
\beq
\chi^i =\frac{\mu}{\sqrt{2\beta}}\,\chi(\rho)\left(\begin{array}{c}0\\0\\1\end{array}\right)\,,
\label{13p}
\eeq
with the boundary conditions  
\beq
\chi(\infty)=0\,,\quad  \chi(0)\approx1\,.
\label{14p}
\eeq
Since the Lagrangian is O(3) invariant any nontrivial solution (\ref{13p}) will introdice two rotational moduli.

Minimizing the energy functional we derive the system of equations for the profile functions
\beqn
\left(\frac{f^{\prime}}{\rho}\right)^\prime&=&a\frac{\varphi^2\,f}{\rho} \,,\nonumber\\[2mm]
\left( \rule{0mm}{4mm} \phi^{\prime}\rho\right)^\prime &=&
\frac{f^2\,\varphi}{\rho}+\frac{\rho\varphi\left(\varphi^2-1\right)}{2}+\frac{\rho\,\varphi\,\chi^2}{2\beta}
\frac{b}{c-1}\,,
\nonumber\\[2mm]
\left(\rule{0mm}{4mm} \chi^{\prime}\rho\right)^\prime
&=& \frac{b}{c-1}\rho\chi\left(c\varphi^2+\chi^2-1\right)\,,
\label{15}
\eeqn
where the primes denote differentiation with respect to $\rho$.

Since our purpose is illustrative -- we try to establish the very possibility of coexistence of the Abelian and non-Abelian strings -- we will limit ourselves to a
particularly interesting choice of one of the parameters,
\beq
 a=1\,,\,\,\,\mbox{ i.e.} \,\,\, m_\phi = m_A\,.
\eeq
In the absence of the $\chi$ field this choice would corresponds to critical or Bogomol'nyi-Prasad-Sommerfield (BPS) \cite{BPS} vortex(srting). The tension of such a string is
\beq
T_0 =2\pi v^2\,.
\label{18}
\eeq
In the numerical solution of the equations of motion we set $a=1\,$. The value of $T_0$ quoted in (\ref{18}) is used as  a reference point.

\section{The \boldmath{$\chi=0$} solution}
\label{inst}

First we consider $\chi=0$ and $\varphi = \varphi_0\equiv \varphi_{\rm ANO}$. One can directly check that this is indeed a solution of the set of of equations (\ref{15}). As in \cite{MSY}, we follow Witten \cite{W} to investigate the stability of the solution with regards to small $chi$ fluctuations. To this end we write down a (linearized) equation for the
$\chi$ modes around the ANO solution. 
The mode equation takes the form
\beq
-\psi^{\prime\prime}+
\left( b\,\frac{c\varphi^2_0-1}{c-1}-\frac{1}{4\rho^2}\right)\psi =\epsilon\psi\,,\qquad \psi (\rho )\equiv\chi \,\sqrt{\rho}\,.
\eeq
Foe two representative values of parameters the
numerical solution yields
\beq
\epsilon = \left\{
\begin{array}{l}
0.041 \,\,\,\mbox{at}\,\,\, b = 0.0987\,,\,\, c = 1.17\,,\\[3mm]
0.234 \,\,\,\mbox{at}\,\,\, b = 1.871\,,\,\, c=2\,.
\end{array}
\label{21}
\right.
\eeq
The positivity of $\epsilon$  implies the stability of the $\chi=0$ solution. The tension of the string  was found to be 
\beq
\frac{T_0}{2\pi v^2}=1-O(10^{-7})\,.
\label{ten}
\eeq
The second number on the right-hand side of Eq. (\ref{ten}) represents the accuracy of our numerical computations.

\section{The \boldmath{$\chi\neq0$} solution}

Now we will demonstrate that although the above ANO solution is locally stable, the model at hand supports
a solution with non-Abelian moduli, i.e. with $\chi\neq 0$. 

In the case of $\chi\neq0$ one can find the asymptotic behavior of the profile functions at $\rho\rightarrow\infty$ by linearizing these equations in this limit,
\beq
 f\sim\sqrt{\rho}\,e^{-\rho}\,, \quad \left(1-\varphi\right) \sim\frac{1}{\sqrt{\rho}}\,e^{-\rho}\,,\quad
 \chi\sim\frac{1}{\sqrt{\rho}}\, e^{-\sqrt{b}\rho}\,.
\eeq
Then we integrated Eqs. (\ref{15}) numerically, keeping $a=1$ and varying parameters $\{b,c,\beta\}$. The plots of the profile functions are shown in Fig. 1. One can note a rather low value of the $\chi$ field in the core. 
In order for the $\chi$ field not to be smeared by quantum fluctuations we must additionally impose a
constraint on the parameters
\beq
 \tilde\lambda\ll\frac{\chi^2(0)}{2(c-1)}\,.
 \label{23}
\eeq
Fortunately, this is always possible since the value of $\tilde\lambda$ is in our hands.
The origin of Eq. (\ref{23}) is as follows. The value of the field $\chi$ in the core of the string should be much larger than the mass, otherwise quasiclassical treatment is not applicable (the condensate of the field should contain many quanta). The mass of the $\chi$ field is given in Eq. (\ref{h8}). The normalization of the field given in Eq. (\ref{core}) should be modified, taking into account the results of our numerical calculation for $\chi(0)$. Thus, the above ratio is expressed as follows
\beq
\frac{\chi^2_{\rm core}}{m^2_{\chi}}=\frac{\mu^2}{2\beta}\,\chi^2(0)\,\frac{1}{\gamma(v^2-\mu^2)}\gg1\,,
\eeq
which reduces to Eq. (\ref{23}).

Similarly to the consideration in Sec. \ref{inst}, we determine the lowest eigenvalue of the equation
\beq
-\psi^{\prime\prime}+
\left[\frac{b}{c-1}\left(c\varphi^2_1-1+3\chi_1^2\right)-\frac{1}{4\rho^2}\right]\psi =\epsilon\psi\,,
\eeq
where $\;\varphi_1$,\, and $\;\chi_1$\, are the  solutions presented in Fig. 1.
This is necessary  to check the stability of $\chi\neq 0$ solution with regards to local variations of $\chi$. The results of numerical calculations yield
\beq
\epsilon = \left\{
\begin{array}{l}
0.042 \,\,\,\mbox{at}\,\,\, b = 0.0987\,,\,\,c = 1.17\,,\\[3mm]
0.235 \,\,\,\mbox{at}\,\,\, b = 1.871\,,\,\, c=2\,.
\end{array}
\right.
\eeq
We determined the tension of the non-Abelian string,
\beq
\frac{T_0}{2\pi v^2}=1-O(10^{-7})\,
\eeq
which must  be compared with Eq. (\ref{ten}). We observe the degeneracy of the two strings (with $\chi=0$ and  $\chi\neq0$).
\newpage

\begin{figure}[h!]
 \begin{minipage}{7.5cm}
\includegraphics[width=7cm]{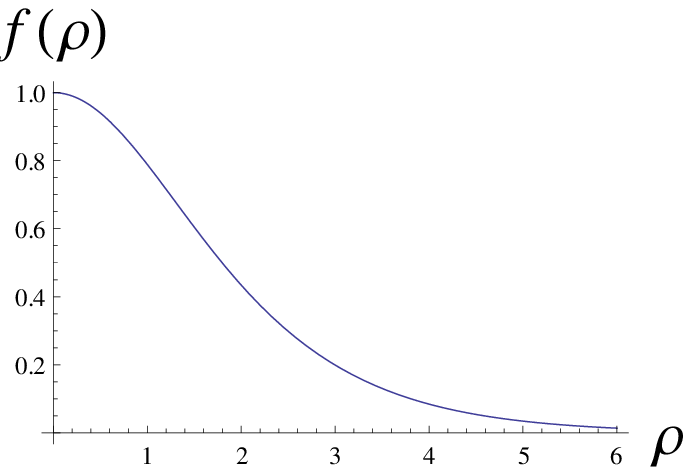}
\includegraphics[width=7cm]{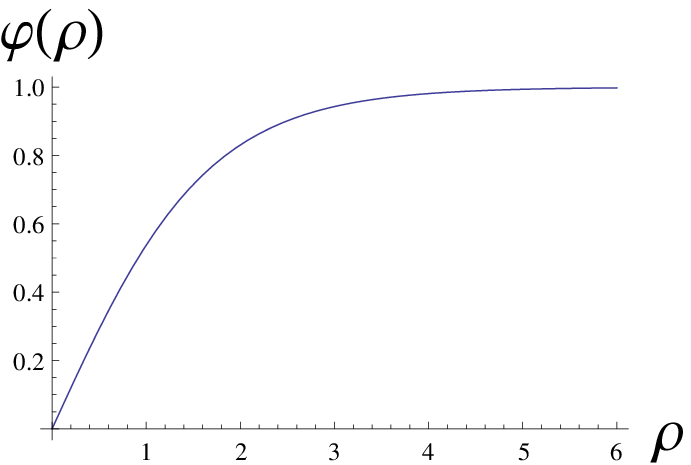}
\includegraphics[width=7cm]{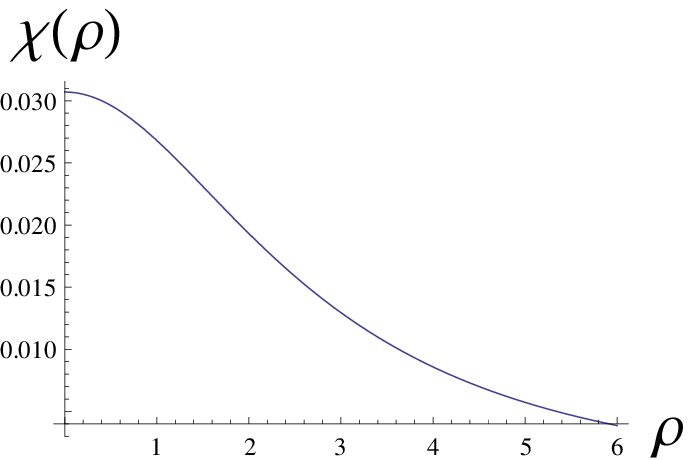}
 \caption{$b=0.0987$, $c=1.17$, $\beta=1.1\beta_*$}
\end{minipage}
\qquad
\begin{minipage}{7.5cm}
\includegraphics[width=7cm]{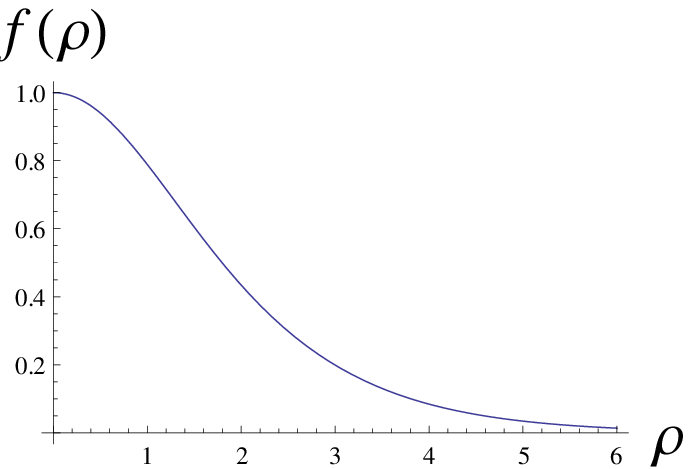}
\includegraphics[width=7cm]{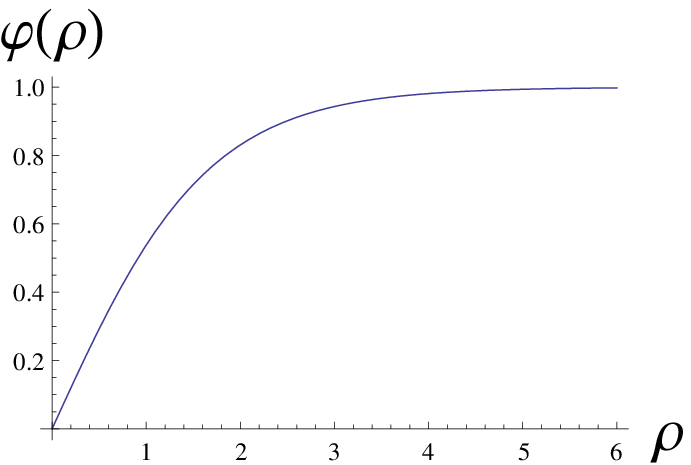}
\includegraphics[width=7cm]{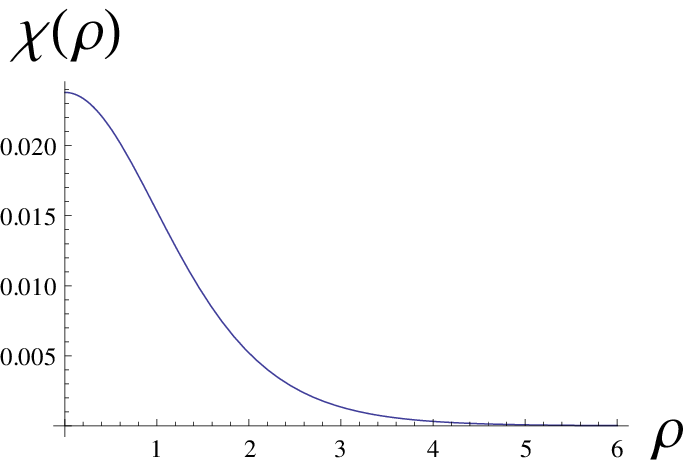}
 \caption{$b=1.871$, $c=2$, $\beta=1.1\beta_*$}
\end{minipage}
\end{figure}
\vspace{2cm}

\section{Conclusions}

In the model \cite{S} we found numerical solutions for the profile functions and calculated the tensions of two distinct (but degenerate) strings. 
This proves the possibility of coexistence of the ANO and non-Abelian degenerate strings in one and the same model simultaneously. The classical degeneracy is not protected against quantum corrections. 
The obvious next step is to supersymmetrize the model to see
whether or not one can have the two strings BPS-saturated. Then the degeneracy will be preserved in higher orders.
Another interesting project is to slightly change the parameters of the model to make the two strings slightly non-degenerate, with the aim of calculating the decay rate of the heavier string into the lighter one.

 \section*{Acknowledgments}

This work  is supported in part by DOE grant DE-FG02-94ER40823.

\end{document}